\documentclass[11pt,a4paper]{article}
\usepackage{cancel, jheppub}

\pdfoutput=1

\usepackage{multirow,xspace}
\usepackage{amsmath,amsfonts,amssymb}
\usepackage{graphicx}
\usepackage[normalem]{ulem}
\usepackage{fancyvrb}
\usepackage{hyperref}
\usepackage{slashed}
\usepackage{booktabs}

\usepackage{cleveref}
\crefname{table}{Table}{Tables}
\crefname{figure}{Fig.}{Figs.}
\crefname{equation}{Eq.}{Eqs.}

\usepackage[usenames,dvipsnames,svgnames,table]{xcolor}
\usepackage{xfrac}
\usepackage{cancel}

\let\oldsqrt\sqrt
\def\sqrt{\mathpalette\DHLhksqrt}
\def\DHLhksqrt#1#2{%
\setbox0=\hbox{$#1\oldsqrt{#2\,}$}\dimen0=\ht0
\advance\dimen0-0.2\ht0
\setbox2=\hbox{\vrule height\ht0 depth -\dimen0}%
{\box0\lower0.4pt\box2}}

\DeclareMathOperator{\SU}{SU}

\DeclareMathOperator{\gU}{U}

\newcommand{\TeV}{\ensuremath{\,\text{Te\kern -0.10em V}}\xspace}
\newcommand{\GeV}{\ensuremath{\,\text{Ge\kern -0.10em V}}\xspace}
\newcommand{\MeV}{\ensuremath{\,\text{Me\kern -0.10em V}}\xspace}

\newcommand{\gev}{\ensuremath{\text{Ge\kern -0.08em V}}\xspace}
\newcommand{\tev}{\ensuremath{\text{Te\kern -0.08em V}}\xspace}

\DeclareRobustCommand{\[}{\begin{equation}}
\DeclareRobustCommand{\]}{\end{equation}}

\title{Electroweak baryogenesis in the $\mathbb{Z}_\mathsf{3}$-invariant NMSSM}
\author[a,c]{Sujeet Akula,}
\author[a,b,c,d]{Csaba Bal\'azs,}
\author[a,e]{Liam Dunn,}
\author[a,b,c]{and Graham White}

\affiliation[a]{School of Physics and Astronomy, Monash University, Victoria 3800, Australia}
\affiliation[b]{Monash Centre for Astrophysics, Monash University, Victoria 3800 Australia}
\affiliation[c]{ARC Centre of Excellence for Particle Physics at the Terascale, Monash University, Victoria 3800 Australia}
\affiliation[d]{Australian Collaboration for Accelerator Science, Monash University, Victoria 3800 Australia}
\affiliation[e]{School of Physics, The University of Melbourne, Melbourne, Victoria 3010, Australia}

\emailAdd{csaba.balazs@monash.edu}
\emailAdd{sujeet.akula@coepp.org.au}
\emailAdd{liamd@student.unimelb.edu.au}
\emailAdd{graham.white@monash.edu}

\abstract{
  We calculate the baryon asymmetry of the Universe in the
  $\mathbb{Z}_\mathsf{3}$-invariant Next-to-Minimal Supersymmetric Standard
  Model where the interactions of the singlino provide the necessary source of
  charge and parity violation.  Using the closed time path formalism, we derive
  and solve transport equations for the cases where the singlet acquires a
  vacuum expectation value (VEV) before and during the electroweak phase
  transition.  We perform a detailed scan to show how the baryon asymmetry
  varies throughout the relevant parameter space. Our results show that the
  case where the singlet acquires a VEV during the electroweak phase transition
  typically generates a larger baryon asymmetry, although we expect that the case 
  where the singlet acquires a VEV first is far more common for any model in which
  parameters unify at a high scale.  Finally, we examine the dependence of the
  baryon asymmetry on the three-body interactions involving gauge singlets.
}

\begin{document}

\maketitle

\section{Introduction}

A cosmological history that includes a period of inflation inevitably washes out
any possible primordial baryon asymmetry in the Universe (BAU). Yet, currently
we observe an asymmetry between baryons and anti-baryons, quantified by the
ratio of the average baryon and entropy densities
\[
  Y_B = \frac{n_B}{s} = \begin{cases}
    8.2\text{--}9.4 \times 10^{-11} ~(95\%~\text{CL}) & \text{BBN~ \cite{Cooke:2013cba}}, \\
    8.65 \pm 0.09 \times 10^{-11} & \text{PLANCK~ \cite{Ade:2015xua}}.
  \end{cases}\label{BAUExp}
\]
Various mechanisms have been proposed to create the observed asymmetry after
inflation and many of these scenarios rely on a thermodynamic phase transition
that may have happened before Big Bang Nucleosynthesis (BBN). The only known
cosmic phase transition that occurred before BBN is the
electroweak phase transition (EWPT). The non-equilibrium condition created
by EWPT is utilised by the electroweak baryogenesis
(EWBG)~\cite{Morrissey:2012db,Cline:2006ts} mechanism to
produce the baryon asymmetry. The presence of the electroweak scale suggests
that tests of EWBG may be within reach
\cite{Riotto:1999yt,Alanne:2016wtx,Long:2017rdo,
White:2016bo,Balazs:2016yvi,Ghorbani:2017jls,Cline:2017qpe,Beniwal:2017eik,Haarr:2016qzq,Vaskonen:2016yiu}.
Aesthetically attractive that EWBG features a common origin for the breaking of both the baryon
and electroweak symmetry.

EWBG requires that the EWPT be strongly first order. In the Standard Model (SM)
the Higgs boson is too heavy to allow for a strongly first order EWPT
\cite{Aad:2015zhl}. The order of the EWPT can be boosted by new weak scale
particles that interact with the Higgs. In supersymmetry, the stops
can catalyse a strongly first order EWPT \cite{Cline:1998hy}.
Search results from the LHC, however, impose severe constraints on such a
possibility
\cite{Huang:2016cjm,Liebler:2015ddv,Laine:2012jy,Francescone:2014pza,Akula:2011zq,Kaufman:2015nda,Nath:2015dza}.
Even if a stop is light enough to boost the order of the EWPT, electric dipole
moment (EDM) constraints render the charge and parity (CP) violating phase
present in the stop-Higgs coupling\footnote{For recent work on CP-violation in the MSSM and other MSSM extensions see \cite{Aboubrahim:2016xuz,Ibrahim:2016rcb}.} to be insufficient~\cite{Berger:2015eba} to
produce the observed baryon asymmetry. These issues in conjunction with the
little hierarchy problem
\cite{Fowlie:2015uga,Okumura:2016yxg,Kim:2013uxa,Farmer:2013eoa,Balazs:2013qva,Akula:2011jx,Akula:2012kk,Akula:2011aa},
which manifests from a combination of the Higgs mass measurement and null
searches for supersymmetric particles, give strong motivation for looking at
extensions to
the minimal supersymmetric scenario\footnote{Especially, when one considers the
motivations for supersymmetry: radiative electroweak symmetry breaking leading
to a light SM-like Higgs, dark matter, solution of the gauge
hierarchy problem, gauge coupling unification in Grand Unified Theories (GUTs),
and string theory.} \cite{Ellwanger:2009dp,Athron:2013ipa}.

Adding a gauge-singlet scalar superfield to the superpotential introduces extra
degrees of freedom that couple to the Higgs, which can boost the strength of the
phase transition \cite{Kozaczuk:2014kva,Balazs:2016tbi,Profumo:2014opa}
and relax the need for the stop mass to be close the EWPT scale\footnote{A
singlet can also ease the little hierarchy problem
\cite{Kim:2013uxa, Fowlie:2014faa}.}. Additionally, experimental constraints on
gauge singlets are not very onerous
\cite{Barger:2007im,Contino:2016spe} and there are strong motivations for
considering gauge singlets beyond baryogenesis. In the case where there is
a discrete $\mathbb{Z}_3$ symmetry between the singlet and Higgs sector, the
Next-to-Minimal Supersymmetric Standard Model (NMSSM) is a proposed solution to
the $\mu $ problem \cite{Fayet:1974pd, Nilles:1982dy, Frere:1983ag,
Derendinger:1983bz, Greene:1986th, Ellis:1986mq, Durand:1988rg, Drees:1988fc,
Ellis:1988er, Pandita:1993tg, Pandita:1993hx, Ellwanger:1993xa, King:1995vk}.
Further, gauge singlets naturally arise in GUTs \cite{Athron:2007en} as well as
in string theory \cite{Li:2006xb}. Moreover, the NMSSM can simultaneously
accommodate inflation, baryogenesis and dark matter \cite{Balazs:2013cia}.
Finally, the singlet can serve to boost the variation of $\beta$ (where
$\tan\beta$ is the ratio of the two Higgs VEVs) during the phase transition by
an order of magnitude compared to the MSSM \cite{Kozaczuk:2014kva}. Since the
BAU is proportional to $\Delta \beta$ there is a possibility that CP violating
phases can be smaller, further evading EDM constraints, and still produce enough
BAU.

The production of the BAU has been explored in the NMSSM via the WKB
approach\footnote{For a derivation of the WKB approach from first principles
see \cite{Kainulainen:2002th,Kainulainen:2001cn}.} \cite{Huber:2006wf}. This
approximation, however, can miss substantial ``memory effects'' that can result
in resonant enhancements of the BAU of up to several orders of magnitude when
the masses are near degenerate\footnote{Recent work \cite{Cirigliano:2009yt}
using Wigner functionals for a toy model suggested
the interesting possibility that the resonance might be severely dampened when
the masses are exactly degenerate. A more precise treatment in
\cite{Cirigliano:2011di} however seemed to indicate there was indeed a
resonance.} \cite{Lee:2004we}. Non-equilibrium quantum field theory has
also been used \cite{Cheung:2012pg,Bian:2017wfv}, utilizing the closed time path
(CTP) formalism \cite{Schwinger:1960qe, Mahanthappa:1962ex, Bakshi:1963bn,
Keldysh:1964ud, Craig:1968XX, Chou:1984es} to calculate the BAU in the presence
of the resonances. Ref.~\cite{Cheung:2012pg}, however, invokes the fast rate
approximation, which can differ from more precise methods by two orders of
magnitude~\cite{Cirigliano:2006wh} in its determination of the BAU.

In this work we derive the transport equations for the most relevant particle
species in the NMSSM for the cases where the singlet acquires a VEV before and
during the electroweak phase transition. We then solve them without assuming
that three body Yukawa, triscalar, strong sphaleron or supergauge interactions
in the Higgses and Higgsino sector are large using the semi-analytic methods
described in Refs.~\cite{White:2015bva,Akula:2016gpl}. We also seek to
answer the question as to whether three body interactions involving gauge
singlets can in principle compete with resonant relaxation terms arising from
CP-conserving interactions with the bubble wall. In the MSSM, such an effect
provides a counter-intuitive boost to the BAU by several orders of magnitude
near resonance despite these three body rates being relaxation terms. Since we are principally concerned with the plausibility of EWBG within the NMSSM
being driven by interactions with the singlino, we root our results in
constraints on the Higgs mass and LHC searches, performing a scan of the relevant
parameter space. Finally we look at how different types of phase transitions can
affect the baryon asymmetry. Whether the singlet acquires a VEV before or during
the electroweak phase transition have a large impact on the determination of the
BAU: if the singlet phase transition occurs before EWPT, then outside of the
bubble, the singlino and Higgsinos have non-zero masses, without thermal
corrections. This affects not only the CP-violating source terms and
CP-conserving relaxation terms, but the thermal decay widths of the singlino.
In fact, the two scenarios change the structure of the transport equations as
fluctuations around the VEV are real by definition (and therefore possess no
asymmetry).

The structure of this paper is as follows. In \cref{sec:NMSSM-intro} we review
the NMSSM and its motivations with a particular emphasis on the structure of the
effective potential. \Cref{sec:transport} contains our derivation of the coupled
transport equations for both phase transition structures mentioned above
outlining our assumptions. In \cref{sec:phasetransition} we discuss the input
parameters of the model with particular emphasis on the thermal widths. We then
solve these equations in \cref{sec:solution} scanning over the NMSSM parameter
space. We examine the effect of both cases: the singlet acquiring a VEV before
or during the EWPT.  We also examine the effect of the three body gauge singlet
transport coefficients in \cref{sec:results}, before a final discussion and
conclusion in \cref{sec:discussion}.

\section{The Next-to-Minimal Supersymmetric Standard Model
\label{sec:NMSSM-intro}}

We calculate the baryon asymmetry within the scale invariant,
$\mathbb{Z}_3$-conserving, NMSSM defined by~\cite{Ellwanger:2009dp} \[ W^{\rm
NMSSM} = W^{\rm MSSM}|_{\mu=0} + \lambda \widehat{S} \widehat{\bf H}_u
\widehat{\bf H}_d + \frac{1}{3} \kappa \hat{S}^3 \text{ .} \] The MSSM
superpotential $W^{\rm MSSM}|_{\mu=0}$, less the $\mu$ term, is defined
in~\cite{Balazs:2009su}, and $\widehat{\bf H}_{u,d}$ and $\widehat{S}$ are
$\SU(2)$ doublet and singlet Higgs superfields, respectively.
In addition to the MSSM soft supersymmetry breaking terms (with $B$ set to zero
\cite{Balazs:2009su}), the scalar potential contains \[ V_{\rm soft}^{\rm NMSSM}
= m_s^2 |S|^2 - \lambda A_\lambda S {\bf H}_u {\bf H}_d + \frac{1}{3} \kappa
A_\kappa S^3 + \text{h.c.}~, \] where ${\bf H}_{u,d}$ and $S$ are the scalar
components of the Higgs doublet (singlet) superfields.  The neutral components
of these scalars acquire a non-zero VEV during electroweak symmetry breaking.
Relative to the MSSM, the new terms in the Higgs potential are \[ V_{\rm H}^{\rm
NMSSM} = \lambda^2 |S|^2 (|{\bf H}_u|^2 + |{\bf H}_d|^2) \lambda^2 |{\bf H}_u
{\bf H}_d|^2 + \kappa^2 |S|^4 + \kappa \lambda S^2 {\bf H}_u^*{\bf H}_d^* +
\text{h.c.}~.  \] To boost the baryon asymmetry we assume complex values for
$\lambda , \kappa ,A_\lambda$ and $A_\kappa$.  Once both the Higgs doublets and
the singlet scalar have acquired a vacuum expectation value, one can write three
re-phasing invariants that appear in the tree level potential
\cite{Cheung:2010ba,Cheung:2012pg} \[
  \phi_\lambda - \phi_\kappa , \quad \phi_\lambda + \phi_{A_\kappa} , \quad
  \phi_\lambda + \phi_{A_\lambda} \ .
\]
Here $\phi_x$ is the complex phase associated with parameter $x\in (\lambda,
\kappa, A_\kappa, A_\lambda )$.  One can then use the CP-odd tadpole conditions
to write two of the rephasing invariants in terms of $\phi_\lambda -\phi _\kappa
$. Therefore there is in reality only a single independent rephasing invariant \cite{Cheung:2011wn}. 
For more details on the NMSSM see Refs.~\cite{Ellwanger:2009dp, Balazs:2009su,
Balazs:2013cia}.


\section{Transport equations\label{sec:transport}}

In this section we derive the set of coupled transport equations that govern the
behaviour of number densities throughout the phase transition. In the NMSSM
there are two possible phase histories that qualitatively change the transport
equations\footnote{This is a reduced list from the four types given in
\cite{Funakubo:2004ka} as this partition is more convenient when discussing
electroweak `baryogenesis.}
\begin{itemize}
  \item Singlet first phase transition (SFPT) phase transition: The singlet
    acquires a VEV before the EWPT.
  \item Singlet spontaneous phase transition (SSPT): The singlet acquires a VEV
    during the EWPT.
\end{itemize}
We make use of the closed time path formalism~\cite{Martin:1959jp,
Schwinger:1960qe, Keldysh:1964ud, Chou:1984es, Mahanthappa:1962ex} following the
procedure given in~\cite{Lee:2004we} to derive the set of transport coefficients
and CP violating source terms.  We will use the usual VEV insertion
approximation (VIA) which assumes that the physics responsible for producing the
BAU is dominated by the region in front of the bubble wall where the Higgs VEV
is small compared to the nucleation temperature and the relevant mass
differences.  The NMSSM includes a resonant source of CP violation in addition
to the CP violating interactions present in the MSSM due to singlino-Higgsino
interactions with the space time varying vacuum.

We ignore first and second generation quarks and squarks as well as all three
generations of leptons and sleptons, invoking the assumption that the rates that
connect these particle species to the rest of the transport equations are small
due to the fact that the relevant Yukawa couplings are small. This is an
assumption that can in some parts of the parameter space be too
rough~\cite{Chung:2009cb} but we leave a thorough investigation of this to
future work.

It has been demonstrated that assuming local equilibrium between third
generation quarks and squarks holds well for large parts of the parameter
space~\cite{Chung:2009qs}. On the other hand, one has to be cautious in assuming
supergauge equilibrium between the Higgs and the Higgsino. Fast supergauge
interactions in the Higgs sector would lead to a suppression of the combination
$\mu_{H_i} - \mu_{\tilde{H}}$ whereas three body interactions involving
singlets/singlinos lead to the suppression of the combination $\mu_{H_i}  + \mu
_{\tilde{H}}$. If one works in the approximation that both types of rates are
fast enough to set the combinations of chemical potentials given before to zero,
then the baryon asymmetry vanishes. While the competition of these types of
rates could indeed suppress the baryon asymmetry we take the precaution of
including both rates in the transport equations rather than assuming they are
large enough to result in local equilibrium relations.

As usual, number densities are defined $n\equiv n-\bar{n}$ where $\bar{n}$ is
the anti-particle density. When the gauge singlet acquires a vacuum expectation
value the fluctuations around this VEV are real and cannot hold any asymmetry.
Therefore in a SFPT phase transition the number density of the singlet is zero.
This means that there is one less transport equation for SFPT phase transitions.
Finally we note that one cannot form a vector charge for the singlino, so we do
not include a number density for the singlino\footnote{In principle one can
define an axial charge as was done here for binos and $\tilde{W}^3$, but the
contribution from doing so was found to be small \cite{Konstandin:2004gy}.}.

Under these simplifying assumptions we are able to derive a set of coupled
transport equations for six charge densities for SSPT phase transitions and five
for SFPT phase transitions as the number density of the singlet $n_S \equiv
n_S-\bar{n}_S$ is zero. We present the more complicated SSPT case as the SFPT
case can be derived from it by setting $n_S$ to zero and modifying transport
coefficients that depend on $v_S$. For a SFPT phase transition, the six linear
combination of number densities which make up the transport equations are
\begin{align}\begin{split}
  n_{H_1} &= n_{H_u^+} + n_{H_u^0} , \\
  n_{H_2} &= n_{H_d^-} + n_{H_d^0} , \\
  n_{\tilde{H}} &= n_{\tilde{H}_u^+} + n_{\tilde{H}_u^0} - n_{\tilde{H}_d^-} -
    n_{\tilde{H}_d^0} , \\
  n_t &= n_{t_R} + n_{\tilde{t}_R} , \\
  n_Q &= n_{t_{L}} + n_{b_{L}} + n_{\tilde{t}_R} + n_{\tilde{b}_L} , \\
  n_S &= n_S .\label{composite}
\end{split}\end{align}
The transport coefficients are derived using the Schwinger-Dyson equations in
the closed time path formalism to relate divergences of current densities to
functions of self energies. These functions of self energies can be expanded in
the chemical potentials of the particles involved in each self energy
interaction. We can relate chemical potentials to number densities in the usual
way. Ignoring terms of $O(\mu^3)$ we can derive the
relation~\cite{Kramers:1940zz} \[ \mu_x = \frac{6}{T^2} \frac{n_x}{k_x} , \]
with \[
  k_x =k_x(0) \frac{c_{F,B}}{\pi^2} \int_{m/T}^\infty \mathrm{d}y\,y
  \frac{e^y}{(e^y\pm 1)^2} \sqrt{y^2-m_x^2/T^2} ,
\]
where $c_{F(B)}=6(3)$ and the sign in the denominator is $\pm$ for fermions and
bosons, respectively. The factors $k_i(0)$ are $2$ for Dirac fermions and
complex scalars and $1$ for chiral fermions. The $k$ factors of our composite
number densities in \cref{composite} are the sum of the $k$ factors for each
component. We then define the linear combinations of rates that act as
coefficients of these composite number densities.

Tree level interactions with space time varying VEVs have CP conserving
components known as ``mass terms'' typically denoted by $\Gamma_m,^{(\cdot )}$
where the superscript describes the particles involved in the interaction
starting with the ``in'' state.  The full set of relevant mass terms
are\footnote{These mass terms typically come in two flavours $\Gamma_m^\pm$
however the negative type is typically much larger so the positive type we
ignore.}
\begin{align}\begin{split}
  \Gamma_m^t &= \Gamma_{m}^{t_R,t_L} + \Gamma_{m}^{\tilde{t}} , \\
  \Gamma_m^{\tilde{H}} &= \Gamma_{m}^{\tilde{H}\tilde{W}^+}
    + \Gamma_{m}^{\tilde{H}\tilde{W}^0}
    + \Gamma_{m}^{\tilde{H}\tilde{B}}
    + \Gamma_{m}^{\tilde{H}\tilde{S}} , \\
  \Gamma_{m}^{H_u H_d} &= \Gamma_{m}^{H_u H_d} , \\
  \Gamma_{m}^{H_1 S} &= \Gamma_{m}^{H_1 S} , \\
  \Gamma_{m}^{H_2 S} &= \Gamma_{m}^{ H_2 S} , \\
  \Gamma_{m}^{\tilde{H} \tilde{S}} &= \Gamma_{m}^{\tilde{H} \tilde{S}} .
\end{split}\end{align}
Note that Higgsino-Higgsino interactions with a space-time varying singlet VEV
do not contribute as the masses are exactly degenerate by definition and the
resonance vanishes when masses are exactly degenerate. Tri-scalar and Yukawa
interactions have a general form related to the functions $I_{F/B}(m_1,m_2,m_3)$
which are defined in Ref. \cite{Cirigliano:2006wh} and for completeness are
also given in \cref{sec:app}.

We define the composite transport coefficients which make up our transport
equations as
\begin{align}\begin{split}
  \Gamma_Y^{tQH_1} &= \frac{12 N_C y_t^2}{T^2} \left[
    c^2 {\cal I}_F(m_{t_R}, m_Q, m_{H_1})
    + |s \mu e^{-i \phi _\lambda }+cA_t|^2
      {\cal I}_B(m_{\tilde{t}_R}, m_{\tilde{Q}}, m_{H_1}) \right] , \\
  \Gamma_Y^{tQH_2} &= \frac{12 N_C y_t^2}{T^2} \left[
    s^2 {\cal I}_F(m_{t_R}, m_Q, m_{H_2})
    + |c \mu e^{-i \phi_\lambda} - sA_t|^2
    {\cal I}_B(m_{\tilde{t}_R}, m_{\tilde{Q}}, m_{H_2}) \right] , \\
  \Gamma_Y^{tQ\tilde{H}} &= \frac{12 N_C y_t^2}{T^2} \left[
    {\cal I}_F(m_{\tilde{H}}, m_Q, m_{\tilde{t}_R})
    + {\cal I}_F(m_{t_R}, m_{\tilde{H}}, m_{\tilde{Q}} \right] , \\
  \Gamma_Y^{H_1, H_2, S} &=  \frac{12|\lambda A_{\lambda}|^2}{T^2}
    {\cal I}_B(m_{H_1}, m_{H_2}, m_{S}) , \\
  \Gamma_Y^{\tilde{H} H_1 \tilde{S}} &= \frac{12 |\lambda|^2}{T^2}
    {\cal I}_F(m_{\tilde{H}}, m_{\tilde{S}}, m_{H_1}) , \\
  \Gamma_Y^{\tilde{H} H_2 \tilde{S}} &= \frac{12 |\lambda|^2}{T^2}
    {\cal I}_F(m_{\tilde{H}}, m_{\tilde{S}}, m_{H_2}) , \\
  \Gamma_Y^{\tilde{H} \tilde{H} S} &= \frac{12 |\lambda|^2}{T^2}
    {\cal I}_F(m_{\tilde{H}}, m_{\tilde{H}},m_S) ,
\end{split}\end{align}
where $c \equiv \cos \alpha$, $s \equiv \sin \alpha$  and $\alpha $ is the Higgs
mixing angle in the unbroken phase.  We also need to include estimates of the
four body scattering terms which in general are small but may become the
dominant contribution when three body decays are kinematically suppressed. For
all cases our four body scattering rate is given by $0.19 |x|^2T/(6 \pi)$ where
$x\in \left\{g_3, \lambda, \lambda A_\lambda\right\}$ is the appropriate
coupling constant. In some regions of parameter space the BAU will be sensitive
to the precise value of these four body scattering rates so future work should
consider a full numerical treatment of these coefficients.

Our transport equations can then be shown to have the form
\begin{align}\begin{split}
  \partial_\mu t^\mu &=
    - \Gamma_{m}^t{\cal U}^m_{t}
    - \Gamma_{Y}^{tQH_1} {\cal U}^Y_{tQH_1}
    - \Gamma_{Y}^{tQH_2} {\cal U}^Y_{tQH_2}
    - \Gamma_{Y}^{tQ\tilde{H}}{\cal U}^Y_{tQ \tilde{H}}
    + \Gamma_{SS} {\cal U}_5
    + S^{\cancel{CP}}_{\tilde{t}}
  ,\\[3pt]
  \partial_\mu Q^\mu &=
      \Gamma_{m}^t{\cal U}^m_{t}
    + \Gamma_{Y}^{tQH_1} {\cal U}^Y_ {tQH_1}
    + \Gamma_{Y}^{tQH_2} {\cal U}^Y_{tQH_2}
    + \Gamma_{Y}^{tQ\tilde{H}} {\cal U}^Y_{tQ \tilde{H}}
    - 2\Gamma_{SS} {\cal U}_5
    - S^{\cancel{CP}}_{\tilde{t}}
  ,\\[3pt]
  \partial_\mu H_1^\mu &=
    - \Gamma_{m}^{H_1H_2} {\cal U}^m _{H_1H_2}
    - \Gamma_{m}^{H_1S} {\cal U}^m_{H_1S}
    + \Gamma_{Y}^{tQH_1}{\cal U}^Y_{tqH_1}
    \\&\qquad\qquad
    - \Gamma_Y^{H_1H_2S}{\cal U}_{H_1H_2S}^Y
    - \Gamma_Y ^{\tilde{H}H_1\tilde{S}} {\cal U}^Y_{\tilde{H} H_1 \tilde{S}}
    - \Gamma_{\tilde{V}}^{H_1\tilde{H}} {\cal U}_{\tilde{V}_1}
  ,\\[3pt]
  \partial_\mu H_2^\mu &=
    - \Gamma_{m}^{H_1H_2} {\cal U}^m_{H_1H_2}
    - \Gamma_{m}^{H_2S}{\cal U}^m_{H_2S}
    + \Gamma_{Y}^{tQH_2} {\cal U}^Y_{tQH_2}
    \\&\qquad\qquad
    - \Gamma_Y^{H_1H_2S}{\cal U}^Y_{H_1H_2S}
    - \Gamma_Y ^{\tilde{H}H_2 \tilde{S}} {\cal U}^Y_{\tilde{H} H_2 \tilde{S}}
    - \Gamma_{\tilde{V}}^{H_2\tilde{H}} {\cal U}_{\tilde{V}_2}
    ,\\[3pt]
  \partial_\mu \tilde{H}^\mu &=
    - \Gamma_{m}^{\tilde{H}} {\cal U}^m_{\tilde{H}}
    - \Gamma_Y^{\tilde{H}S} {\cal U}^Y_{\tilde{H}S}
    + \Gamma_Y^{tQ \tilde{H}} {\cal U}^Y_{t Q \tilde{H}}
    - \Gamma_Y^{\tilde{H} H_1 \tilde{S}}{\cal U}^Y_{\tilde{H}H_1\tilde{S}}
    \\&\qquad\qquad
    - \Gamma_Y^{\tilde{H} H_2 \tilde{S}}{\cal U}^Y_{\tilde{H} H_2 \tilde{S}}
    + \Gamma_{\tilde{V}}^{H_1 \tilde{H}} {\cal U}_{\tilde{V}_1}
    + \Gamma_{\tilde{V}} ^{H_2 \tilde{H}} {\cal U}_{\tilde{V}_2}
    + S^{\cancel{CP}}_{\tilde{H}\tilde{W}}
    - S^{\cancel{CP}}_{\tilde{H}\tilde{S}}
  ,\\[3pt]
  \partial_\mu S^\mu &=
    - \Gamma_{m}^{H_1S}{\cal U}^m _{H_1S}
    - \Gamma_{m}^{H_2S} {\cal U}^m_{H_2S}
    + \Gamma_Y^{H_1H_2S} {\cal U} ^Y_{H_1H_2S}
    - \Gamma_Y^{\tilde{H}S} {\cal U}^Y_{\tilde{H}S}
  \ .
  \label{NMSSMQTEs}
\end{split}\end{align}
Here we have defined the combinations of chemical potentials as follows
\begin{align}\begin{split}
  {\cal U}^m_{t} &= \left( \frac{t}{k_t} -
  \frac{Q}{k_Q} \right) , \ \ \ {\cal U}^Y_{tQH_i} = \left( \frac{t}{k_t}-
  \frac{Q}{k_Q} - \frac{H_i}{k_{H_i}} \right) , \ \ \ {\cal U}^Y_{tQ \tilde{H}}
  =  \left( \frac{t}{k_t}- \frac{Q}{k_Q} - \frac{\tilde{H}}{k_{\tilde{H}}}
  \right) , \\
  {\cal U}^m_{H_1H_2} &=
  \left(\frac{H_1}{k_{H_1}}+\frac{H_2}{k_{H_2}} \right) , \ \ \ {\cal U}^m
_{H_iS} = \left( \frac{H_i}{k_{H_i}} + \frac{S}{k_S} \right) , \ \ \ {\cal
  U}_{H_1H_2S}^Y = \left(\frac{H_1}{k_{H_1}}+\frac{H_2}{k_{H_2}}+ \frac{S}{k_S}
  \right) , \\
  {\cal U}^Y_{\tilde{H} H_i \tilde{S}} &= \left(
  \frac{\tilde{H}}{k_{\tilde{H}}}+\frac{H_i}{k_{H_i}} \right) , \ \ \ {\cal U}
_{\tilde{V}_i} = \left(\frac{H_i}{k_{H_i}}- \frac{\tilde{H}}{k_{\tilde{H}}}
  \right) , \ \ \ {\cal U}^m_{\tilde{H}}= \frac{\tilde{H}}{k_H} , \ \ \ {\cal U}
^Y_ {\tilde{H}S}  = \left(\frac{S}{k_S}\right) .
\end{split}\end{align}

As usual, the axial chemical potential, ${\cal U}_5$, is given by $\mu_L-
\mu_R$. Since only the left handed quark doublet and the right handed top is
sourced, this reduces to \[
  {\cal U}_5 = \frac{2Q}{k_Q}- \frac{t}{k_t}+\frac{9(Q+t)}{k_b} \text{ .}
\]
Finally, the strong sphaleron rate is taken to be \(\Gamma_{SS} \approx 16
\alpha_S^4 T\)~\cite{Moore:2000mx}.

\section{Thermal parameters\label{sec:phasetransition}}

The production of the BAU is resonantly enhanced when the masses of the singlino
and Higgsino (calculated in the symmetric phase) are nearly degenerate. The
width and height of the resonance, and therefore the width of the allowed
parameter space, are controlled by the thermal widths of the singlino and the
Higgsino. The magnitude of the thermal widths for various particles in the NMSSM
is typically dominated by gauge interactions and are proportional to the
involved coupling constant squared. This means that the thermal widths for
(s)quarks tend to be quite large. However, in the absence of gauge interactions,
the thermal width of the singlino is expected to be small. Indeed the only two
places where the thermal width of the singlino or the singlet can be broadened
is through Yukawa interactions.

Let us divert more attention to the thermal width of the singlino as the thermal
width of the singlet only weakly affects the BAU. The relevant Yukawa
interactions involve the singlino, Higgsino and the Higgs with relevant
couplings of $\kappa$ and $\lambda$. From Ref.  \cite{Elmfors:1998hh} the
thermal width that results from such Yukawa interactions is \[
  \Gamma_{\tilde{S}} = b(m_H^*) 0.01 (\lambda^2 + \kappa^2) ,
\]
where $b(m_H^*)$ is a function which monotonically decreases with the mass of
the Higgs in the symmetric phase. For a SSPT phase transition the Higgs mass is
just the Debye mass under the VIA, where we assume the physics primarily
responsible for the production of the BAU occurs in the symmetric phase just
ahead of the bubble wall. The Higgs mass in the symmetric phase of a SFPT phase
transition gets contributions from the singlet VEV and is thus boosted.  The
thermal width of the singlino is therefore much smaller for a SFPT phase
transition.

We estimate the remaining thermal widths, diffusion coefficients and bubble wall
properties in \cref{parameters}.  For the diffusion coefficients we use the
values given in~\cite{Joyce:1994zn} from which we also derive the rest of our
thermal widths. For the parameter $\Delta \beta $ we note that a feature of the
NMSSM is it can be an order of magnitude larger than its MSSM
value~\cite{Kozaczuk:2014kva}. However, since the BAU is linearly proportional
to $\Delta \beta$ we just take the fiducial value of $0.05$
along with a fiducial value of the CP violating phase which we set to its
maximal value. Although the value of $\Delta \beta$ can be an order of magnitude
greater in the NMSSM compared to the MSSM \cite{Kozaczuk:2014kva}, the BAU is
linearly proportional to $\Delta \beta$ so it is simple to translate our results
to the case where $\Delta \beta$ is large. Since we assume that $\Delta \beta$
is small we assume that the mixing angle in the symmetric phase - that is near
the bubble wall - is equal to its zero temperature value evaluated at the $Z$
boson mass. The BAU tends to get larger for small values of the bubble wall
velocity which can also vary over an order of magnitude \cite{Kozaczuk:2014kva}.
We take a moderate value.  Finally, our thermal mass from the singlino agrees
with \cite{Cheung:2012pg} and is given by \[
  \Delta_T m_{\tilde{S}}^2 = \frac{|\lambda|^2 +2 |\kappa |^2}{8}T^2\ .
\]

\begin{table}[t]
  \centering
  \begin{tabular}{@{}ll@{}} \toprule
    Parameter & Value \\ \midrule
    $D_Q$ & $6/T$ \\
    $D_{\tilde{H}} $ & $110/T$ \\
    $D_t$ & $6/T$  \\
  $D_{S} $ & $150/T$ \\ \bottomrule
  \end{tabular}
  \hspace{2pt}
  \begin{tabular}{@{}ll@{}} \toprule
    Parameter & Value \\ \midrule
    $D_{H_u}$ & $110/T$ \\
    $D_{H_d}$ & $110/T$ \\
    $\Gamma_{\tilde{t}}$ & $0.5T$ \\
    $\Gamma_{\tilde{B}}$ & $0.04T$ \\ \bottomrule
  \end{tabular}
  \hspace{2pt}
  \begin{tabular}{@{}ll@{}} \toprule
    Parameter & Value \\ \midrule
    $\Gamma_{H_{u,d}}$ & $0.025T$\\
    $\Gamma_{\tilde{W}} $ & $0.065T$ \\
    $\Gamma_{\tilde{H}} $ & $0.025T$ \\
    $\Gamma_S$ & $0.003T$ \\ \bottomrule
  \end{tabular}
  \hspace{2pt}
  \begin{tabular}{@{}ll@{}} \toprule
    Parameter & Value \\ \midrule
    $\Delta \beta $ & $0.05$ \\
    $v_W$ & $0.05$ \\
    $L_W $ & $5/T$ \\
    $T_N$ & $100$ \\ \bottomrule
  \end{tabular}
  \caption{\label{parameters}
    The base set of parameters used for our numerical study. The diffusion
    constants are taken from \cite{Joyce:1994zn}. Here $a=\{1,0.5,0.25\}$ for
    the case where the mass of the singlet is $\in
    [0,0.25T],[0.25T,0.5T],[0.5T,\infty]$, respectively, and $b=\kappa ^2 +
    \lambda^2$.
  }
\end{table}

\section{Semi-analytic solution \label{sec:solution}}

Neglecting the bubble wall curvature we can reduce the problem to a single
dimension by solving the system in the rest frame of the bubble wall
$z=|v_Wt-x|$. We then use the diffusion approximation to write $\vec{\nabla}
\cdot \vec{J} = \nabla^2 n$ thus reducing the problem to a set of coupled
differential equations in a single space time variable. To answer the question
as to whether the singlino can drive the production of the baryon asymmetry we
set all CP violating phases apart from the singlino-Higgsino-Higgs interaction
to zero.  Transport equations in this form have a closed form analytical
solution in each phase \cite{White:2015bva}.

Consider the case where there is no high temperature singlet VEV. In the broken
phase the solution is \[
  X(z) = \sum_{i=1}^{12} x_1 A_X(\alpha_i) e^{-\alpha_i z}
    \left(\int_0^z \mathrm{d}y\, e^{-\alpha_i y}
      S_{\tilde{S}}^{\cancel{CP}}(y) - \beta_i \right) \text{ ,}
\]
and in the symmetric phase we have \[
  X(z) = \sum_{i=1}^{12} A_{X,s}(\gamma_i) y_i e^{\gamma_i z} \text{ ,}
\]
where, \[ X \in \left\{Q, t, H_1, H_2, \tilde{H}, S, \right\}\text{ .}\] The
derivation of $\alpha_i, \beta_i, x_i, y_i, \gamma_i$ and
$A_{X,(s)}(\alpha_i(\gamma_i))$ is given in~\cite{White:2015bva}. From these
solutions one can then define the left handed number density \(n_L(z) = Q_{1L} +
Q_{2L} + Q_{3L} = 5Q + 4T\).  The baryon number density, $\rho_B$, satisfies the
equation~\cite{Carena:2002ss, Cline:2000nw} \[
  D_Q \rho_B''(z) - v_W \rho_B'(z) - \Theta(-z) {\cal R} \rho_B
  = \Theta(-z) \frac{n_F}{2} \Gamma_{ws} n_L(z) ,
\]
where $n_F$ is the number of fermion families. The relaxation parameter is given
by \[
  {\cal R} = \Gamma_{ws} \left[\frac94 \left(1 + \frac{n_{sq}}{6}\right) +
  \frac32\right] \ ,
\]
where $n_{sq}$ is the number of thermally available squarks and \(\Gamma_{ws}
\approx 120\,\alpha_W^5 T\)~\cite{Bodeker:1999gx, Moore:1999fs, Moore:2000mx}.
The baryon asymmetry of the Universe, $Y_B$ is then given by \[
  Y_B = -\frac{n_F \Gamma_{ws}}{2\kappa_+ D_Q S} \int_{-\infty}^0
    e^{-\kappa_ - x}\, n_L(x) \,\mathrm{d}x \label{BAU} ,
\]
where \[ \kappa_\pm = \frac{v_W \pm \sqrt{v_W^2 + 4 D_Q {\cal R}}}{2 D_Q} , \]
and the entropy is \[ S = \frac{2 \pi ^2}{45} g_* T^3 .  \]


\section{Numerical results\label{sec:results}}

We sample the NMSSM parameter space using \texttt{MultiNest v3.10} \cite{Feroz:2008xx, Feroz:2007kg, Feroz:2013hea} and calculate the spectrum with \texttt{
SOFTSUSY v3.7.2} \cite{Allanach:2001kg,Allanach:2013kza} by restricting the Higgs mass to $125\GeV$ rather than performing a global fit. The parameter space is reduced to a few dimensions by
fixing the values of the soft masses for second and third third generation
sfermions to $6\TeV$, third generation to $3\TeV$ and $M_1$ to $500\GeV$.  The
prior distributions in the remaining free parameters are given in
\cref{tab:parameter_ranges}.  The parameters $m_{H_u}$, $m_{H_d}$, $m_{S}$ are
set by tadpole conditions and $A_\kappa$ is set to get the right Higgs mass.
The range of parameters are chosen with the following considerations in mind
\begin{enumerate}
  \item Since we are considering only the CPV source that is not present in the
    MSSM we require the singlino and Higgsino masses (including thermal
    corrections) to be relatively close before EWSB and not very heavy compared
    to a plausible value of the nucleation temperature. This latter concern is
    to avoid severe Boltzmann suppression of the CPV source and the former
    concern is to ensure a resonant enhancement of the CPV source.
  \item The singlet mass cannot be too heavy compared to the Higgs since it must
    catalyse a strongly first order EWPT (we assume that the stop is
    too heavy to perform such a role).
  \item Converse to the previous consideration, a light singlet with large
    mixing with the standard model Higgs will be ruled out by collider
    constraints.
\end{enumerate}

The VEV insertion approximation leads us to take the masses and degrees of
freedom in the symmetric phase as it is in this phase where the total left
handed number density biases unsuppressed electroweak sphalerons producing the
baryon asymmetry \cite{Lee:2004we}.  We sanitise the results by removing points
where the VEV insertion approximation is unreliable --- that is, where the mass
gap between the singlino or Higgsino mass (including the Debye mass) is large
enough to spuriously change the sign of CP conserving relaxation terms. We also
note that when the mass gap between the singlino and the Higgsino is very small
the VEV insertion may become reliable \cite{Cirigliano:2009yt,Cirigliano:2011di}.

\begin{figure}[t]
  \centering
  \includegraphics[height=9cm]{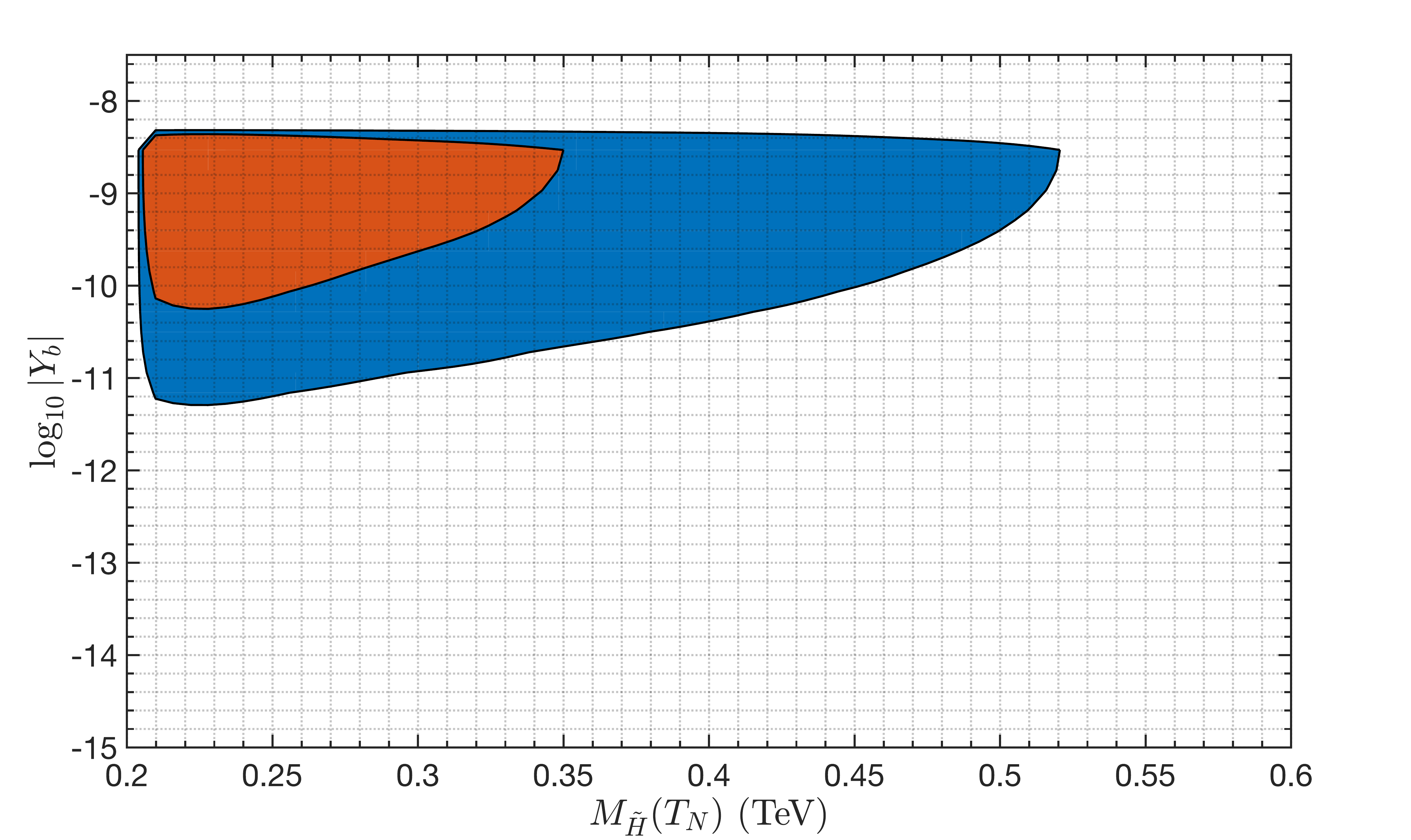}
  \includegraphics[height=9cm]{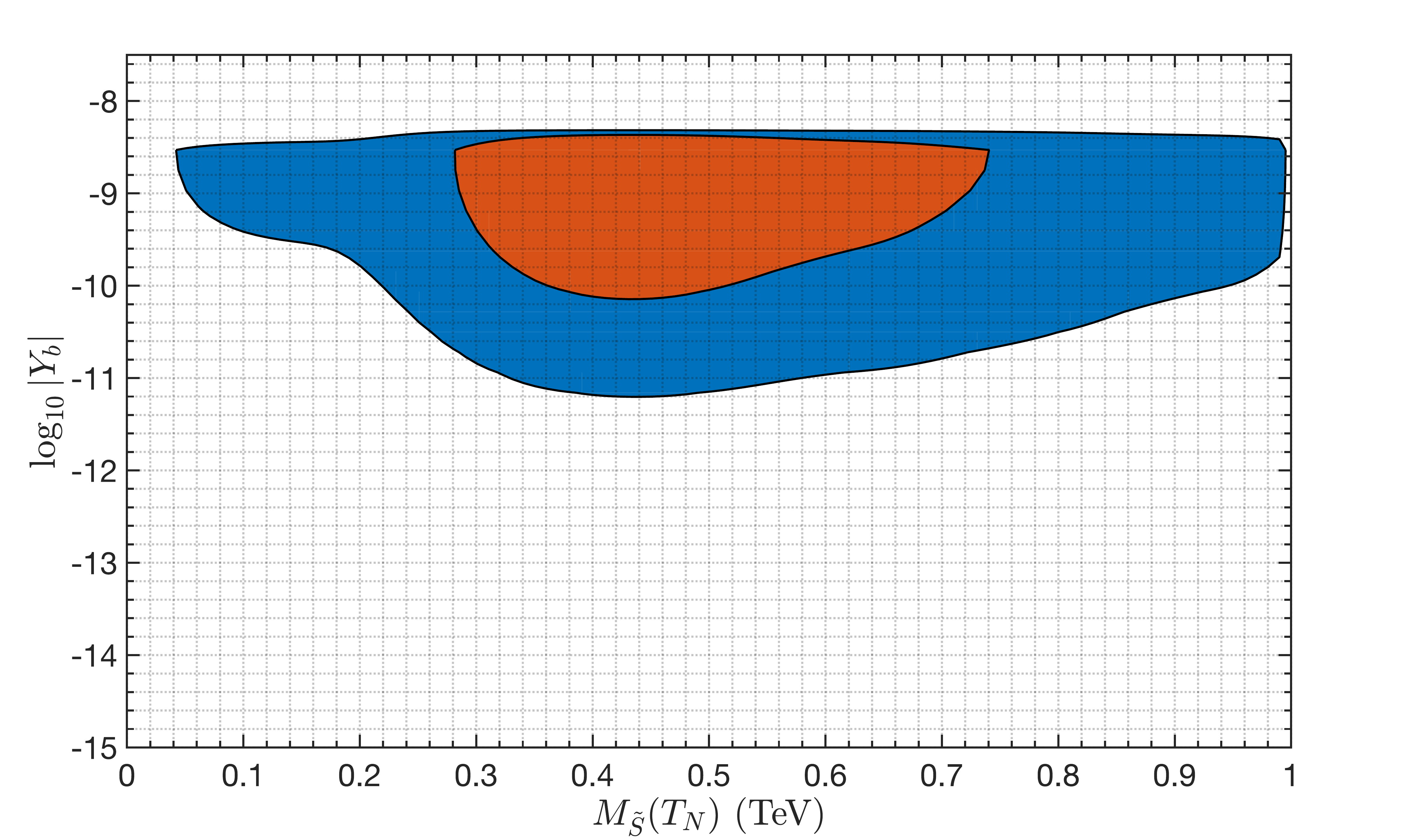}
  \caption{ \label{fig:SSPT}
    Two dimensional posterior probability distribution in the baryon asymmetry,
    produced by the maximum CP-violating phase in the SSPT (singlet simultaneous) scenario, and the
    thermal mass of the Higgsino (top panel) and singlino (bottom panel).
    We colour the $1\sigma$ and $2\sigma$ credible regions in orange and blue, 
    respectively.
  }
\end{figure}

\begin{figure}[t]
  \centering
  \includegraphics[height=9cm]{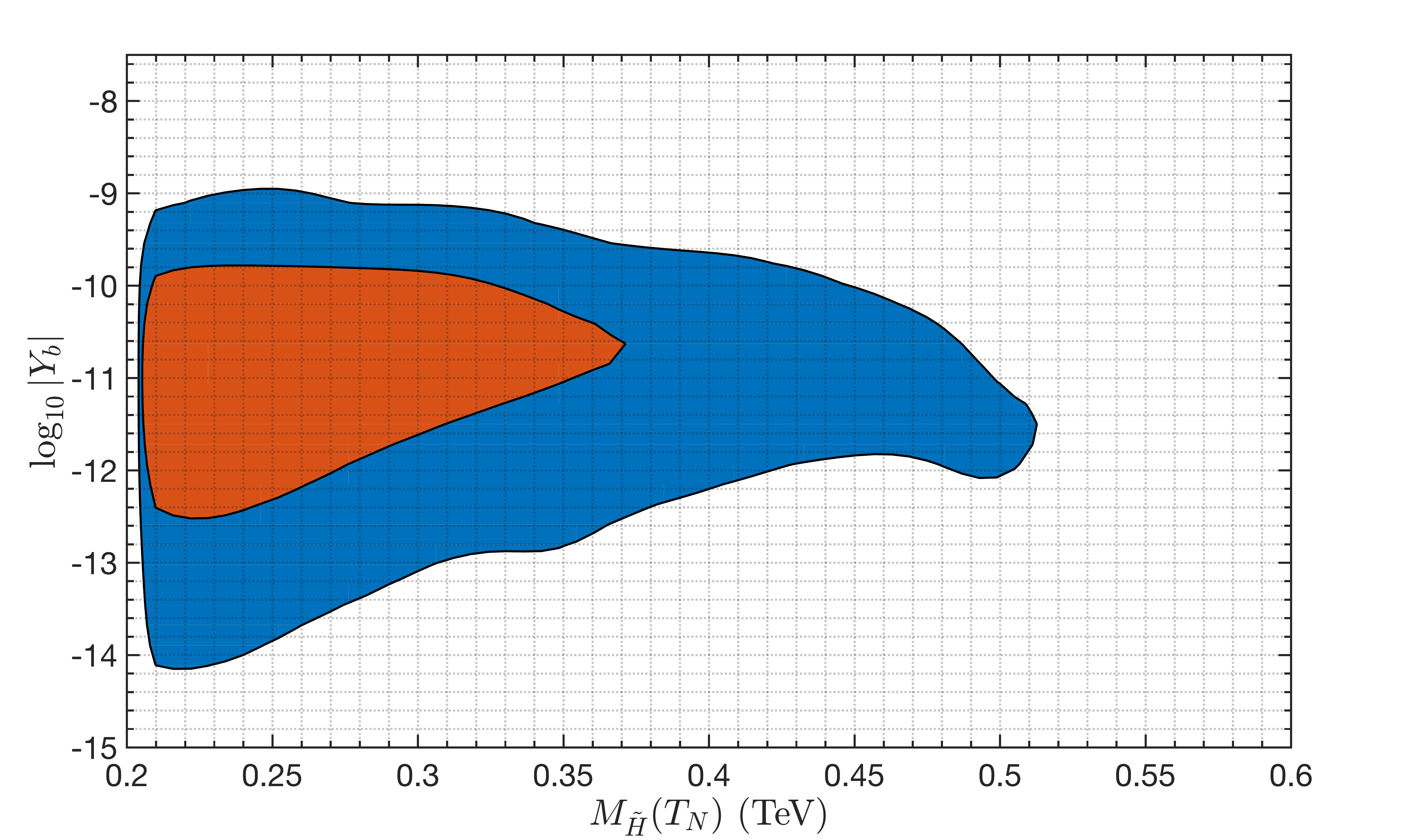}
  \includegraphics[height=9cm]{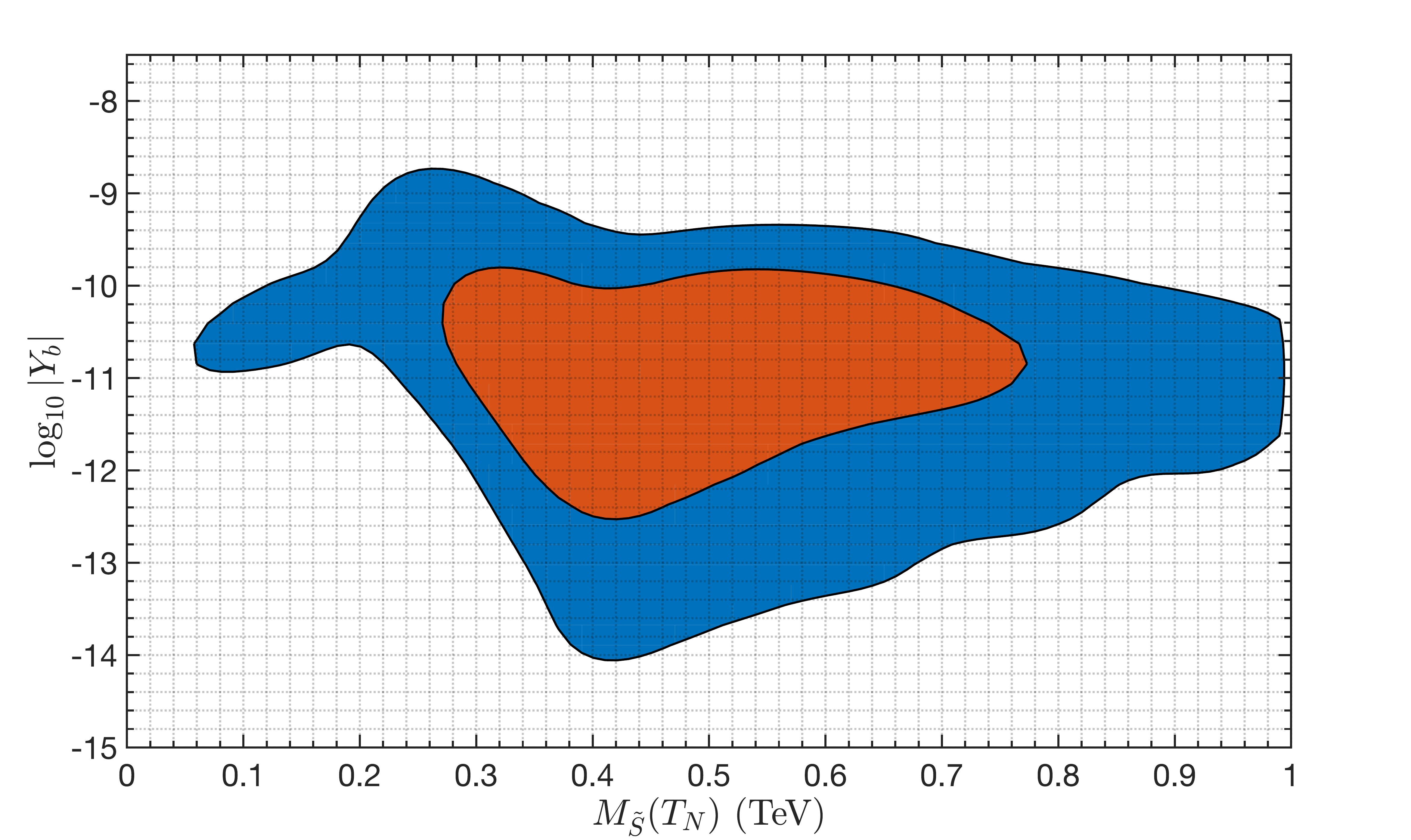}
  \caption{ \label{fig:massplots}
    Two dimensional posterior probability distribution in the baryon asymmetry,
    produced by the maximum CP-violating phase in the SFPT scenario, and the
    thermal mass of the Higgsino (top panel) and singlino (bottom panel).
    We colour the $1\sigma$ and $2\sigma$ credible regions in orange and blue, 
    respectively.
  }
\end{figure}

We also sanitise all other mass relaxation terms (e.g. $\Gamma_m^{\tilde{t}}$)
by setting them to a random positive infinitesimal number when the in- and
virtual-state are far from degeneracy and the naive calculation of the rate
yields a negative number. This avoids the spurious case where the mass
relaxation terms change sign rather than decaying to zero due to the breakdown
of the VEV insertion approximation and can give a spurious boost the the
BAU\footnote{A random infinitesimal number is chosen rather than zero for the
sake of the stability of our code but there is no discernible numerical
difference in the BAU between setting these values to zero or a small number.}.
We set the CP violating phase to its maximal value, $\sin \phi =1$, as the BAU
is linearly proportional to the sin of this phase.  The BAU is also proportional to $\Delta
\beta$ which we set to a value of $0.05$ 
as it can be it can have a range of $\sim [0.01,0.2]$ in the NMSSM
\cite{Kozaczuk:2014kva}. Any point in our scan that has a BAU greater than the
observed value for such a CPV phase and value of $\Delta \beta$ can be interpreted
as a point where the correct value of the CP violating phase is \begin{equation} \frac{\Delta \beta \sin \phi
}{ F} = \frac{Y_{B}^{\rm obs}}{Y_B} \ , \end{equation} 
where $F = 0.05$.
For each parameter point we calculate the BAU for SFPT and SSPT 
if the square of the Higgs mass at $H=S=0$ is positive when one includes
the Debye mass as well as corrections from the thermal functions $J_B$ and
$J_F$.  If the potential does not have a positive curvature at $H=S=0$ for any
reasonable range of nucleation temperature ($\lesssim 200\GeV$) the implication is
that the origin of field space is a local maximum rather than a local minimum.
This means that the phase transition must be either SFPT or second order for
these points in the parameter space. As a GUT scale model (with soft masses on
the order of the electroweak scale in magnitude) can drag $m_{H_u}^2$ to large
negative values when it is evolved to the electroweak scale, it could be harder
to find regions of parameter space that survive this cut; the SFPT
is the more realistic case. This analysis we leave to a future study. 
For our prior ranges the majority of the sample does indeed survive this
cut. For the points that survive this cut we calculate the baryon asymmetry. 
We then calculate the posterior distributions based on the mass of
the Higgs and the criteria that both the singlino and Higgsino masses are less
than a \tev. We colour the $1\sigma$ and $2\sigma$ credible regions in orange and
blue respectively as shown in \cref{fig:SSPT} for the SSPT. We perform a similar
analysis in the SFPT case and find that generically this scenario produces a
lower asymmetry as shown in \cref{fig:massplots}. Indeed the largest value of
the BAU is an order of magnitude large in the SSPT compared to the case of the
SFPT. This suppression is due to the fact that the soft masses of the singlino
and Higgsino are both proportional to the VEV of the singlet which tends to be
quite large.

\begin{table}[t]
  \centering
  \begin{tabular}{ccccccc}
    \toprule
    $A_\lambda$ &  $A_\kappa $ & $M_2$  & $\tan \beta $ & $\lambda $ & $\kappa $ & $ \lambda v_S $ \\
    \midrule
    $[-4000, 4000]$  & $[0,200] $ & $[100,1000] $ & $ [1.1,5] $ & $[0.001,0.5] $ & $ [0.001,0.5]$ & $[200,800] $ \\
    \bottomrule
  \end{tabular}
  \caption{\label{tab:parameter_ranges}
    Sample ranges for the scanned NMSSM parameters. All dimensionful numbers are in \gev.
    $M_1$ was fixed to $500\GeV$. The modest range of $\tan \beta$ was to help
    satisfy flavour constraints. Sfermions were decoupled by setting the first
    and second generation squark masses to $6\TeV$ and the third generation to $3\TeV$.
  }
\end{table}

\begin{figure}[t]
  \centering
  \includegraphics[width=15cm]{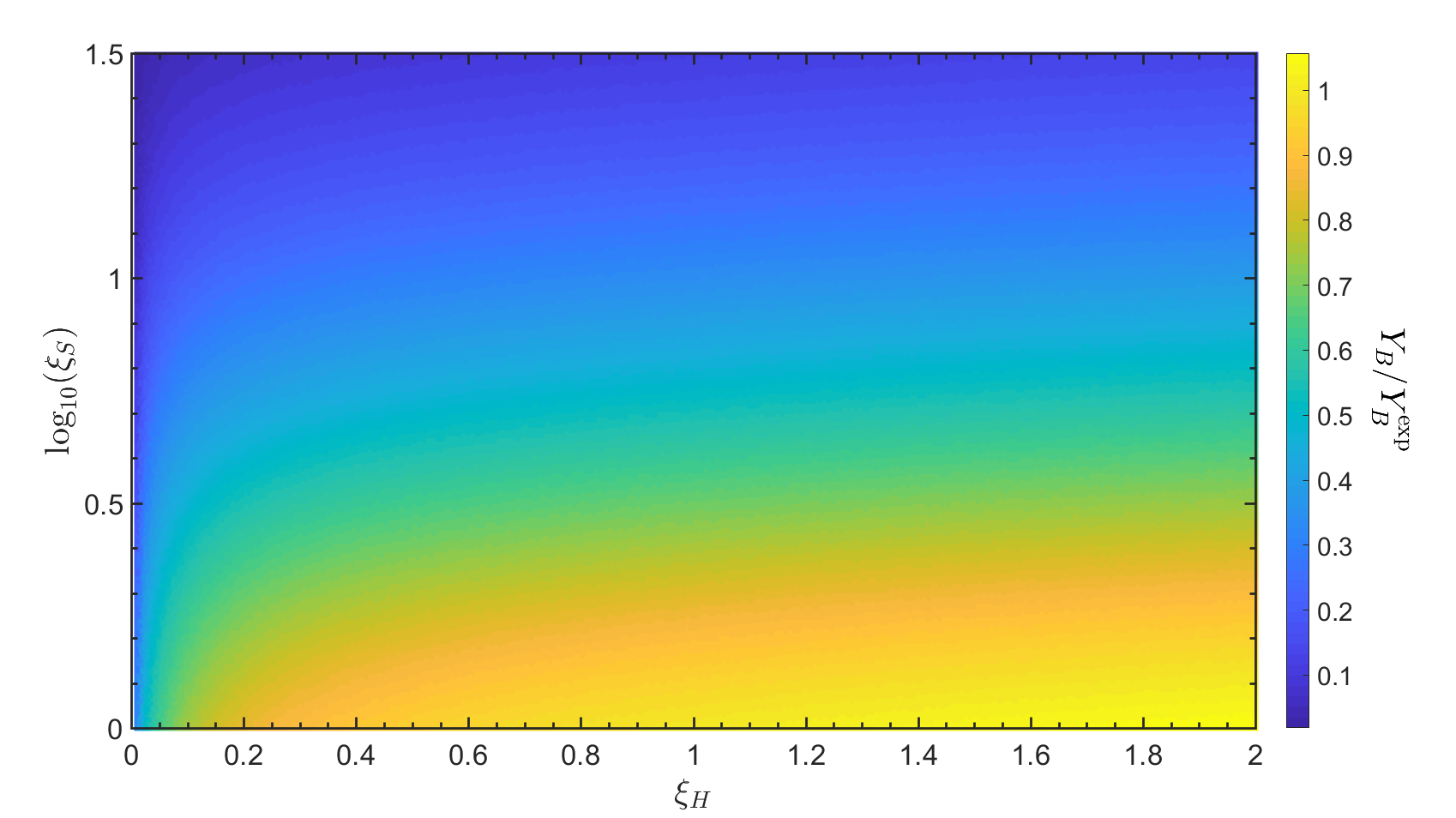}\\
  \includegraphics[width=15cm]{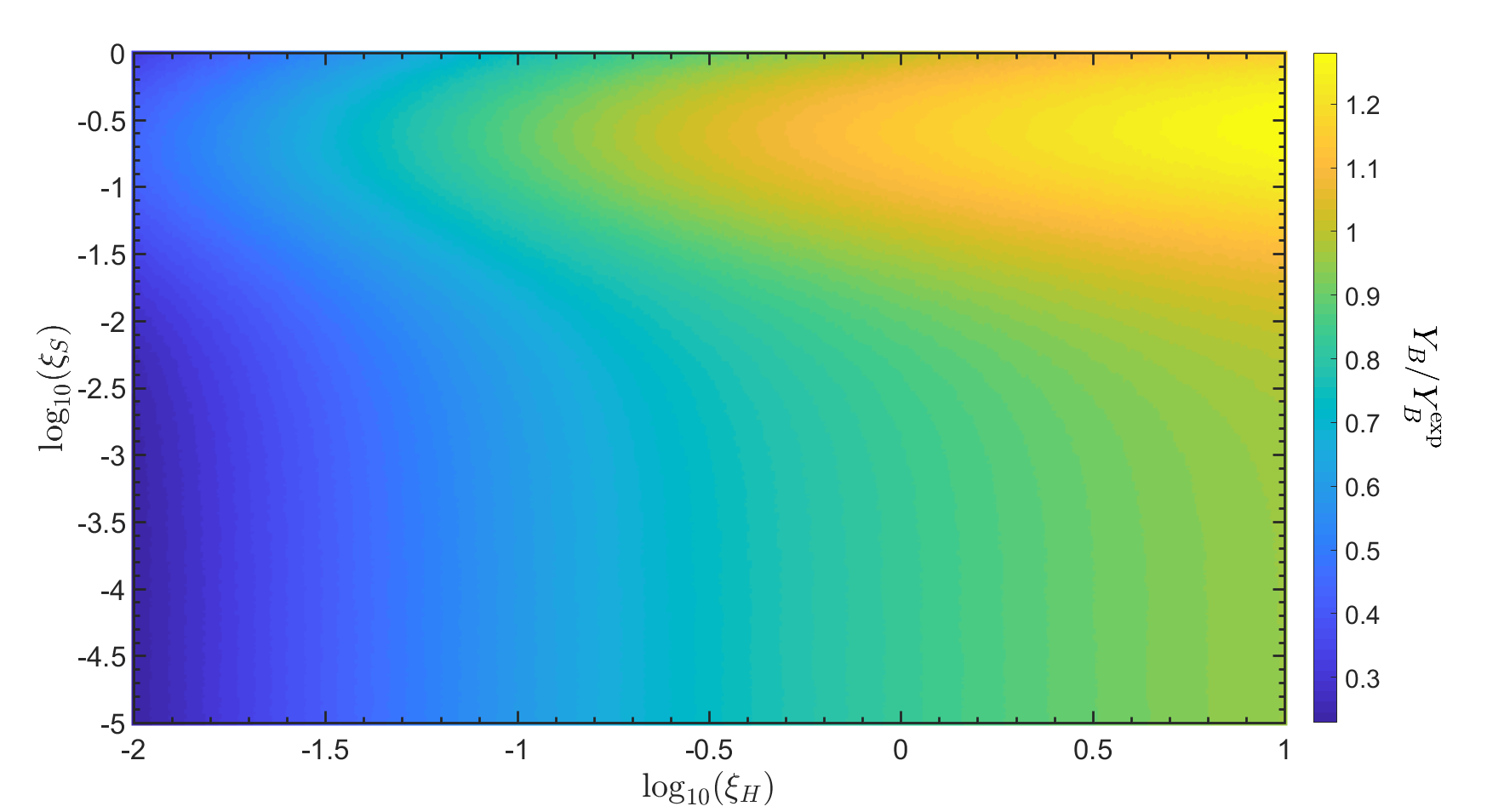}
  \caption{\label{ybap12}
    Baryon asymmetry for a benchmark point with lowest value of $\chi^2$ for a
    SFPT (singlet first) phase transition with varying values of the three
    body rates.  The horizontal axis varies the three body rates involving (s)tops and
    Higgs(inos). The vertical axis varies the three body rates involving singlets or singlinos.
  }
\end{figure}

%

In performing the scan for SFPT we make the approximation that the singlet VEV
does not change throughout the electroweak phase transition. This results in an
underestimate of the baryon asymmetry if the singlet VEV is smaller in the EW
symmetric phase then the EWSB phase. The reason for this is that the VEV of the
singlet contributes to the masses of the particles involved in the CPV source
and their contribution is usually large enough that Boltzmann suppression can
become an issue.  This along with the greater variability of $\Delta \beta$
really motivates future work where the dynamics of the phase transition and the
calculation of the baryon asymmetry are performed simultaneously.
\begin{table}[]
  \centering
  \begin{tabular}{ccccccc}
    \toprule
    $A_\lambda$ &  $A_\kappa $ & $M_2$  & $\tan \beta $ & $\lambda $ & $\kappa $ & $ \lambda v_S $ \\
    \midrule
    $519 $  & $4.80 $ & $977 $ & $ 4.18 $ & $0.39 $ & $ 0.39$ & $249 $ \\
    \bottomrule
  \end{tabular}
  \caption{\label{tab:benchmark}
    Benchmark point with the lowest value of $\chi^2$ for an assumed
    SFPT. Dimensionful parameters are listed in \gev. Note that neither this,
    nor any other point in our sample have had
    its zero temperature phenomenology checked in detail beyond the correct
    masses of standard model particles (including the Higgs) as well as some
    rules of thumb we explain in the text.
  }
\end{table}

\begin{figure}[t]
  \centering
  \includegraphics[width=15cm]{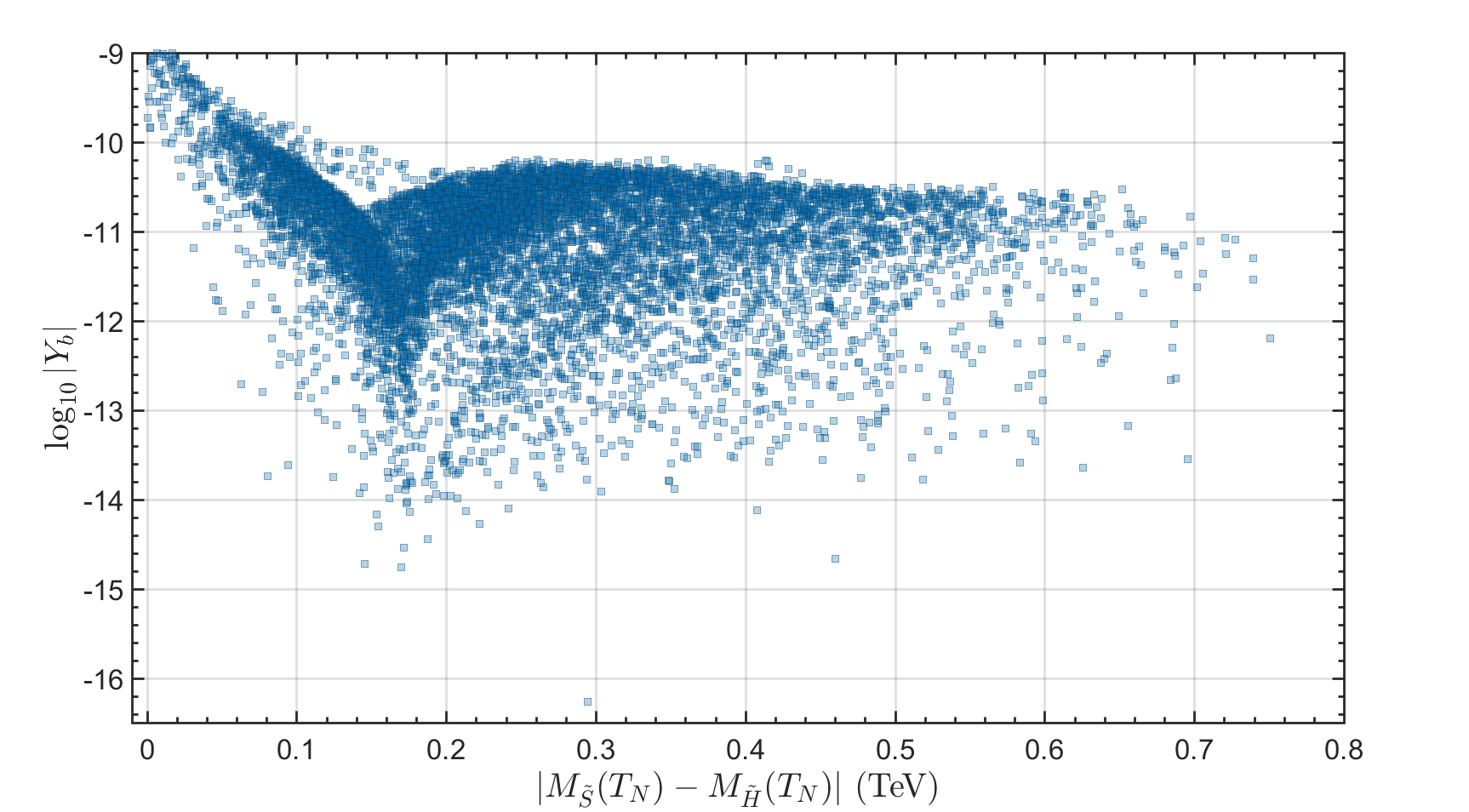}
  \includegraphics[width=15cm]{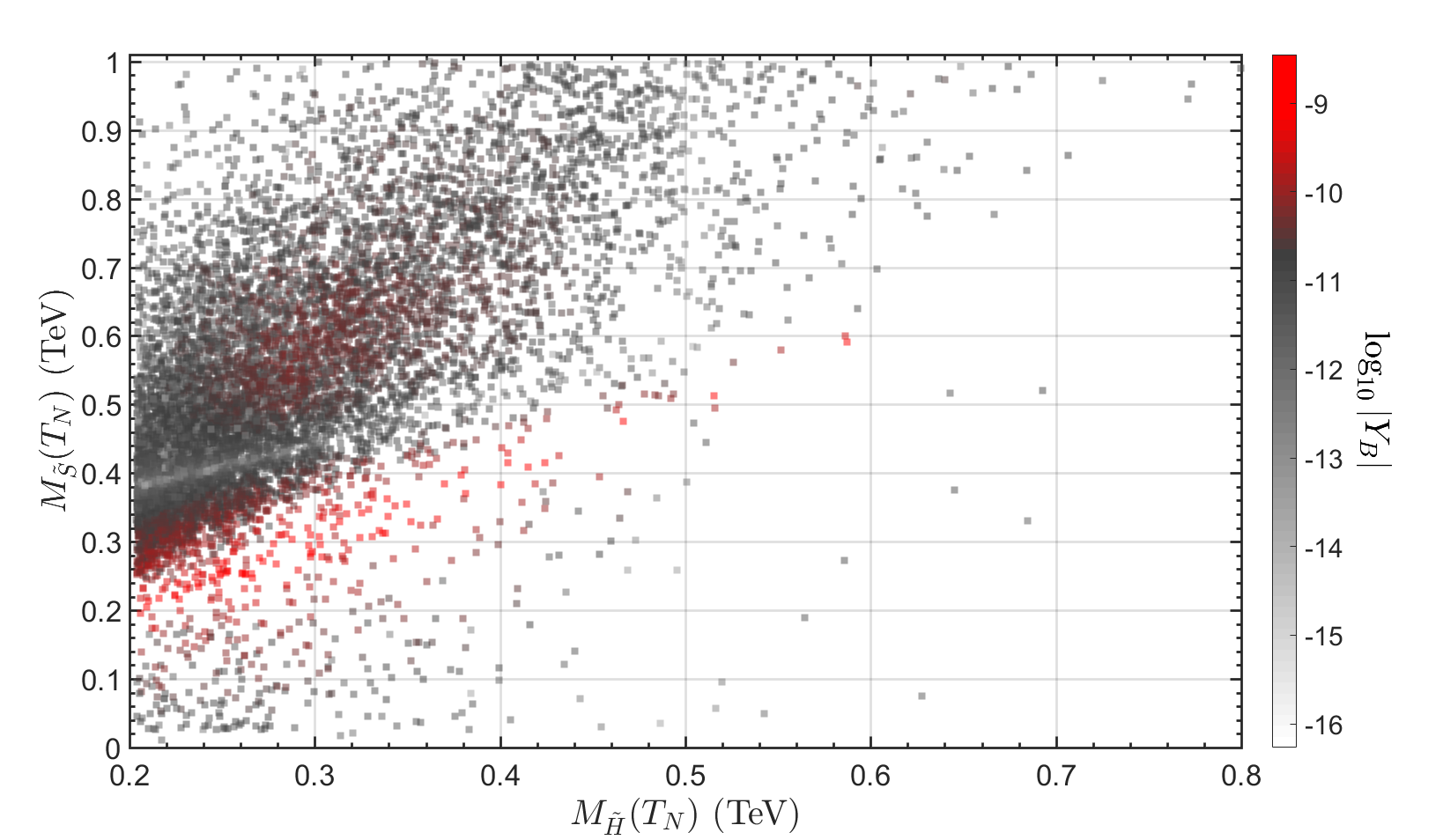}
  \caption{\label{fig:resonanceplots}
    Baryon asymmetry produced during a SFPT against the mass difference of the
    singlino and Higgsino masses including the Debye masses (top panel) and a
    contour plot of the baryon asymmetry against the same aforementioned masses
    (bottom panel). The resonance is the dominant feature of the plot and it
    appears that in the SFPT case, one needs to be near the center of the
    resonance to produce the observed BAU unless $\Delta \beta$ is very large.
    The possibility of large $\Delta \beta$ in the NMSSM leads to the
    interesting possibility of off resonance baryogenesis.
  }
\end{figure}
%

In \cref{fig:massplots} we show the range of zero temperature masses for the
Higgsino and singlino against the BAU for a SFPT phase transition.  There is a
substantial proportion of the parameter space with a sufficiently large BAU.  We
find that the resonant enhancement that occurs when the masses of the singlino
and Higgsino are near degenerate is the dominant predictor of a large BAU. This
is clear from \cref{fig:resonanceplots}. Of particular interest to us is the
fact that well off resonance one can still obtain a BAU that is close to an
order of magnitude below the observed rate. This shortcoming in the BAU, however, 
can be made up for by a sufficiently large $\Delta \beta$. This possibility opens 
up another avenue in which the BAU can be produced within the NMSSM off resonance.
As a caveat we note that as one ventures further off resonance, more 
skepticism should be held toward the accuracy of the result as
the VEV insertion approximation is losing its reliability. 


In Ref. \cite{Cirigliano:2006wh} it was shown that the BAU can vary by orders of
magnitude with the variation of the magnitude of three body rates involving
Higgs(inos) and (s)tops. The NMSSM has new three body rates involving
Higgs(inos) and singlet/singlinos. We show how the BAU varies as a function of
both types of three body rates for a benchmark point which has the lowest value
of $\chi^2$ (given in \cref{tab:benchmark}). We introduce the factors $\xi_H$ and $\xi_S$
to the transport equations given in \cref{NMSSMQTEs} so that every three body
rate $\Gamma_Y$ not involving singlets or singlinos is multiplied by $\xi_H$,
and those involving singlets or singlinos are multiplied by $\xi_S$. E.g., \[
    \Gamma_Y^{tQH_1} \mapsto \xi_H \Gamma_Y^{tQH_1} \text{, ~~ and ~~}
    \Gamma_Y^{\tilde{H}H_1\tilde{S}} \mapsto
    \xi_S \Gamma_Y^{\tilde{H}H_1\tilde{S}}
    \text{ .}
\]
The BAU increases with the three body rates involving (s)tops and Higgs(ino)
interactions. We believe the reasons are the same as that given in
Ref.~\cite{Cirigliano:2006wh}. For three body rates involving singlet (or singlino)
interactions we find a different behaviour. When these rates are very small the
BAU increases to a peak, similar to the 3 body rates involving (s)quarks. However, the
BAU drops sharply after the peak.  We explain the sharp drop by the fact that these three body
rates relax the linear combination $\mu_{\tilde{H}} + \mu_H \approx 0$, whereas
both the supergauge rates as well as all other triscalar and Yukawa rates
conspire to relax the combination $\mu_{\tilde{H}} - \mu_H\approx 0$
\cite{Chung:2009qs}.  The supergauge rate is typically a moderately large value,
so if the singlino-Higgsino rate is also large then the result is that both
$\mu_{\tilde{H}}$ and $\mu_H$  are relaxed to zero. Since the BAU is, in a fast
rate approximation, proportional to $\mu_H$ \cite{Lee:2004we}, the BAU goes to
zero as well.

\section{Discussion and Conclusion\label{sec:discussion}}

Electroweak baryogenesis is an attractive paradigm for producing the BAU due to
its testability. In fact testability is an unavoidable feature of this paradigm
as any physics that is responsible for catalysing the baryon production during
the EWPT must have mass scales at (or just above \cite{Balazs:2016yvi}) the weak
scale and non-trivial couplings. In this paper we have examined one of the most
popular extensions to the standard model - the NMSSM - and indeed we find that
the scenario requires that at least some neutralinos must be relatively close to
the weak scale.  If the singlino and Higgsino are both light, the contribution
to dark matter from a neutralino lsp tends to be smaller than the observed
value. It would certainly be interesting to test whether the electroweak
baryogenesis constraints derived in this paper are compatible with getting the
right dark matter abundance, or whether one needs to extend the NMSSM.  Apart
from the constraints on the parameter space, we also examined the structure of
the transport equations keeping in mind the different possibitites of how the
EWPT proceeds.  As usual the most striking feature is the existence of a
resonant boost in the CP violating source when the masses of the singlino and
Higgsino are near degenerate. In the MSSM one can also have a surprising boost
to the baryon asymmetry in the case where three body rates involving stops and
Higgs are large. We have new three body rates involving the Higgs, Higgsino and
singlino which enhances the BAU up to a peak and then suppresses the baryon
asymmetry when they get large. The suppression is quite severe when its size
becomes large enough to compete with the supergauge rate involving Higgs and
Higgsinos. This we put down to these two interactions creating approximate local
equilibrium relations which conflict to set the baryon asymmetry to
approximately zero.

On SSPT phase transitions we can make the qualitative comment that in general it
is easier to get a resonance boost to the baryon asymmetry in a SSPT phase
transition where the masses of the singlino and the Higgsino are just the Debye
masses in the region just outside the bubble of broken elecroweak phase which is
primarily responsible for the BAU production. There are reasons however to be
skeptical on whether SSPT would frequently occur in any GUT scale model. A full
statistical analysis of this we leave to future work. SSPT phase transitions
also present technical challenges. The masses of the neutralinos in the broken
phase might have a larger contribution from the VEV of the singlet than the
Higgs. So the space-time variation of these masses during the electroweak phase
transition is very large stretching faith in the VEV insertion approximation. A
full Wigner functional treatment as well as a numerical study of the phase
transition would shed further light on the viability of baryogenesis in SSPT
phase transitions.

Finally we conclude by noting that the relationship between masses requiring a
resonance in order to produce enough BAU makes electroweak baryogenesis a fairly
fine tuned mechanism. The NMSSM has the attractive possibility of providing more
paths to such a boost through producing a large enough $\Delta \beta$ which
could potential make EWBG work even well off resonance. A detailed numerical
study of $\Delta \beta$ in the NMSSM would shed light on how realistic this
possibility is.

\acknowledgments{
  The authors thank Peter Athron for many elucidating discussions of the Higgs
  sector of the NMSSM. One of us (S.A.) thanks Angelo Monteux for a helpful
  exposition of cosmological parameter measurements. G.W. acknowledges Michael
  Ramsey-Musolf and
  his group for some helpful discussions of this work, and would also like to
  acknowledge the Australian Postgraduate Award. This work in part was supported
  by the ARC Centre of Excellence for Particle Physics at the Terascale.
}

\appendix

\section{Transport coefficients and sources \label{sec:app}}

For completeness we present the three body rates here. For a triscalar
interaction one has
\begin{multline}
  {\cal I}_B (m_1,m_2,m_3 ) = \frac{1}{16 \pi^3}
    \int_{m_1}^\infty d\omega_1 h_B (\omega_1) \\
  \times \left\{ \log \left(\frac{ e^{\omega_1/T} - e^{\omega_2^+/T} }{
    e^{\omega_1 /T} - e^{\omega_2^-/T} }
  \frac{ e^{\omega_2^-/T} - 1 }{ e^{\omega_2^+/T}-1 } \right) \left[ \Theta (m_1 -m_2
  - m_3)- \Theta (m_2-m_1-m_3) \right] \right. \\ \left.
  + \log \left(\frac{e^{-\omega_1 /T}- e^{\omega_2^+/T}}{e^{-\omega_1 /T}-
  e^{\omega _2^-/T}} \frac{e^{\omega_2^-/T}-1}{e^{\omega_2^+/T}-1} \right)
  \Theta (m_3-m_2-m_1)\right\} ,
\end{multline}
whereas the the three body Yukawa rate is
\begin{multline} {\cal I}_F (m_1,m_2,m_3 ) = -\left( m_1^2+m_2^2-m_3^2
    \right) \frac{1}{16 \pi^3} \int_{m_1}^\infty d\omega_1 h_F (\omega_1)
    \nonumber \\ \times \left\{ \log \left(\frac{e^{\omega_1 /T}+ e^{\omega
_3^+/T}}{e^{\omega_1 /T}+ e^{\omega_3^-/T}} \frac{e^{\omega
_3^-/T}-1}{e^{\omega_3^+/T}-1} \right) \left[ \Theta (m_1 -m_2 - m_3)-
    \Theta (m_3-m_1-m_2) \right] \right. \nonumber \\ \left. + \log
  \left(\frac{e^{-\omega_1 /T}+ e^{\omega_2^+/T}}{e^{-\omega_1 /T}+ e^{\omega
_3^-/T}} \frac{e^{\omega_3^-/T}-1}{e^{\omega_3^+/T}-1} \right) \Theta
  (m_2-m_2-m_3) \right\} ,
\end{multline}
where \[ h_{B/F}(x) = \frac{e^{x/T}}{(e^{x/T}\pm 1)^2} .\]

\clearpage


\providecommand{\href}[2]{#2}\begingroup\raggedright\endgroup

\end{document}